\documentclass[prd,twocolumn,a4paper,superscriptaddress,floatfix]{revtex4}
\usepackage{bbm}
\usepackage[all]{xy}
\usepackage{amsmath}
\usepackage{amssymb}
\usepackage[pdftex]{graphicx}
\usepackage{epstopdf}

\def\be{\begin{equation}}
\def\ee{\end{equation}}
\newcommand{\bes}{\begin{subequations}}
\newcommand{\ees}{\end{subequations}}

\def\ben{\begin{eqnarray}}
\def\een{\end{eqnarray}}
\def\ba{\begin{array}}
\def\ea{\end{array}}

\begin{document}
\newcommand{\half}{{\textstyle\frac{1}{2}}}
\allowdisplaybreaks[3]
\def\a{\alpha}
\def\b{\beta}
\def\g{\gamma}\def\G{\Gamma}
\def\d{\delta}\def\D{\Delta}
\def\ep{\epsilon}
\def\et{\eta}
\def\z{\zeta}
\def\t{\theta}\def\T{\Theta}
\def\l{\lambda}\def\L{\Lambda}
\def\m{\mu}
\def\f{\phi}\def\F{\Phi}
\def\n{\nu}
\def\p{\psi}\def\P{\Psi}
\def\r{\rho}
\def\s{\sigma}\def\S{\Sigma}
\def\ta{\tau}
\def\x{\chi}
\def\o{\omega}\def\O{\Omega}
\def\k{\kappa}
\def\pa {\partial}
\def\ov{\over}
\def\br{\\}
\def\ud{\underline}

\newcommand\lsim{\mathrel{\rlap{\lower4pt\hbox{\hskip1pt$\sim$}}
    \raise1pt\hbox{$<$}}}
\newcommand\gsim{\mathrel{\rlap{\lower4pt\hbox{\hskip1pt$\sim$}}
    \raise1pt\hbox{$>$}}}
\newcommand\esim{\mathrel{\rlap{\raise2pt\hbox{\hskip0pt$\sim$}}
    \lower1pt\hbox{$-$}}}

\title{The role of domain wall junctions in Carter's pentahedral model}

\author{P.P. Avelino}
\email[Electronic address: ]{ppavelin@fc.up.pt}
\affiliation{Centro de F\'{\i}sica do Porto, Rua do Campo Alegre 687, 4169-007 Porto, Portugal}
\affiliation{Departamento de F\'{\i}sica da Faculdade de Ci\^encias
da Universidade do Porto, Rua do Campo Alegre 687, 4169-007 Porto, Portugal}
\author{J.C.R.E. Oliveira} 
\email{jespain@fe.up.pt} 
\affiliation{Centro de F\'{\i}sica do Porto, Rua do Campo Alegre 687, 4169-007 Porto, Portugal} 
\affiliation{Departamento de Engenharia F\'{\i}sica da Faculdade de Engenharia
da Universidade do Porto, Rua Dr. Roberto Frias, s/n, 4200-465 Porto, Portugal}
\author{R. Menezes} 
\email{rms@fisica.ufpb.br} 
\affiliation{Centro de F\'{\i}sica do Porto, Rua do Campo Alegre 687, 4169-007 Porto, Portugal} 
\affiliation{Departamento de F\'\i sica, Universidade Federal da Para\'\i ba, Caixa Postal 5008, 58051-970 Jo\~ao Pessoa, Para\'\i ba, Brasil} 
\author{J. Menezes}  
\email{jmenezes@ect.ufrn.br} 
\affiliation{Centro de F\'{\i}sica do Porto, Rua do Campo Alegre 687, 4169-007 Porto, Portugal} 
\affiliation{Escola de Ci\^encias e Tecnologia, Universidade Federal do Rio Grande do
Norte, Avenida Hermes da Fonseca, 1111, 59014-615 Natal, RN, Brasil}

\begin{abstract}

The role of domain wall junctions in Carter's pentahedral model is investigated both analytically and numerically. We perform, 
for the first time, field theory simulations of such model with various initial conditions. We confirm that there are very specific 
realizations of Carter's model corresponding to square lattice configurations with X-type junctions which could be stable. However, we show that more realistic realizations, consistent with causality constraints, do lead to a scaling domain wall network with Y-type junctions. We determine the network properties and discuss the corresponding cosmological implications, in particular for dark energy.

\end{abstract} 
\maketitle

\section{Introduction}

There is now overwhelming observational evidence that our Universe is presently undergoing an era of accelerated expansion \cite{Komatsu:2008hk,Frieman:2008sn}. In the context of general relativity such period can only be explained if the universe is permeated with an exotic dark energy component violating the strong energy condition. The dark energy is often described  by a nearly homogeneous scalar field minimally coupled to the other matter fields. If the scalar field is static then it is equivalent to a cosmological constant but the more interesting case is definitely that of a dynamical scalar field \cite{Copeland:2006wr}.

Nevertheless, the dark energy role is not necessarily played by a (nearly) homogeneous field. In fact, it has been claimed that a frozen domain wall network could naturally explain the observed acceleration of the universe \cite{Bucher:1998mh}. However, this possibility has been seriously challenged by recent observational results which favor a dark energy equation of state parameter, $w$, very close to $-1$ (note that $w=-2/3+v^2 \ge -2/3$ for domain walls, where $v$ is the root mean square velocity). Furthermore,  although it is possible to build (by hand) stable domain wall lattices there is strong analytical and numerical evidence that no such lattices will ever emerge from realistic phase transitions \cite{PinaAvelino:2006ia,Avelino:2006xy,Avelino:2006xf,Battye:2006pf,Avelino:2008ve}. This provided strong support for a no-frustration conjecture invalidating domain walls as a viable dark energy candidate.

Still, it has been argued that winding domain wall models with X-type junctions could give rise to static lattice type configurations thus accounting for at least a fraction of the dark energy density \cite{Carter:2004dk,Battye:2005hw,Battye:2005ik,Carter:2006cf}. Carter's pentahedral model \cite{Carter:2004dk,Carter:2006cf} has been constructed as an example of a model having an odd number of vacuum configurations giving rise to an even type system through the formation of X-type junctions. However, in \cite{PinaAvelino:2006ia,Avelino:2006xy} the claim that Carter's pentahedral model would form X-type junctions has been challenged and it was argued that Y-type junctions would be formed instead. In this paper we definitely settle this question.

Throughout the paper we use units in which $c=\hbar=m=1$, where the mass scale, $m$, can be chosen arbitrarily.

\section{The model}

Consider the action
\begin{equation}\label{eq:L}
S=\int d^4x \, \sqrt{-g} \mathcal \, {\cal L} \, ,
\end{equation}
where $\Phi$ and $\Psi$ are complex scalar fields, 
\begin{equation}
{\cal L}=X+Y-V\,,
\end{equation}
$V(\Phi,\Psi)$ is the scalar field potential,
\begin{eqnarray}
X&=&\frac{1}{2} {\Phi^*}_{,\mu} \Phi^{,\mu} \,,\\
Y&=&\frac{1}{2} {\Psi^*}_{,\mu} \Psi^{,\mu}\,,
\end{eqnarray}
and the superscript $^*$ stands for the complex conjugate. Carter's pentahedral model has a potential given by \cite{Carter:2004dk,Carter:2006cf}
\begin{eqnarray}
V&=&V_0\left(\left( |\Phi |^2-1\right)^2+\left( |\Psi|^2-1\right)^2\right) \nonumber\\
&+&V_\epsilon  \left(|\Phi |^2 |\Psi|^2 \left(\cos \theta+\cos \chi \right)+2/(1-\epsilon)\right)\,,
\label{eq:mex}
\end{eqnarray}
where $\Phi=|\Phi|\, e^{i\phi}$, $\Psi=|\Psi|\, e^{i\psi}$, $0 < \epsilon < 1$, $V_\epsilon = \epsilon V_0$, $\theta=2 \phi + \psi$ and $\chi=2 \psi-\phi$. The potential has five 
minima with $V=0$ satisfying $\cos \theta=\cos \chi=-1$ and $|\Phi|^2=|\Psi|^2=1/(1-\epsilon)$. 

If $\epsilon = 0$ then Eq.~(\ref {eq:mex})  represents the standard Mexican hat potential. Hence, the model allows for 
cosmic string solutions associated with regions where the phase of $\Phi$ and/or $\Psi$ changes by $2 n \pi$, where 
$n$ is an integer. The string width is roughly $\delta_s \sim V_0^{-1/2}$ and consequently its energy per unit length 
is given by
\begin{equation}
\mu \sim \delta_s^2 V_0 \sim 1\,.
\end{equation}

For $\epsilon \ll 1$  the static domain wall trajectories connecting different minima may be calculated to first order in $\epsilon$ assuming that $|\Phi|=|\Psi|=1$ 
everywhere. In this case 
\begin{eqnarray}
{\cal L}&=&\frac{1}{2} {\phi}_{,\mu} \phi^{,\mu} + \frac{1}{2} {\psi}_{,\mu} \psi^{,\mu} - V(\phi,\psi) \,,\nonumber\\
&=& \frac{1}{10} {\theta}_{,\mu} \theta^{,\mu} + \frac{1}{10} {\chi}_{,\mu} \chi^{,\mu} - V(\theta,\chi)\,,
\end{eqnarray}
with the potential given approximately by
\begin{eqnarray}
V(\theta,\chi) &=& V_\epsilon  \left(\cos \theta+\cos \chi +2\right)\,\nonumber\\
&=& 2 V_\epsilon  \left(\cos^2 (\theta/2)+\cos^2 (\chi/2)\right)\,.
\end{eqnarray}
Assuming that $|\Phi|=|\Psi|=1$ everywhere would be enough to guarantee that Y-type junctions never occur, as long as the energy density remains finite everywhere. If the above condition is relaxed then Y-type junctions are no longer forbidden and 
will be associated with cosmic strings ($|\Phi| \sim |\Psi| \sim 0$ on the string core). 

\begin{figure}[t!]
\hspace{-0.3cm}\includegraphics[width=4.0cm,height=4.0cm]{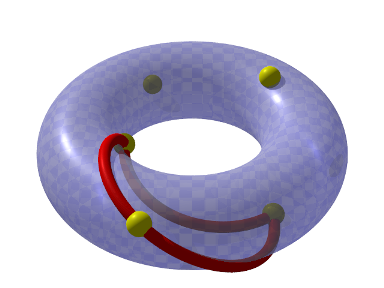}\hspace{-0.3cm}\includegraphics[width=4.0cm,height=4.0cm]{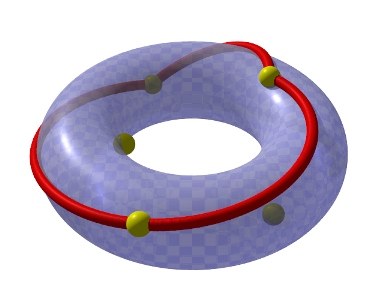}
\\
\vspace{-0.2cm}
\hspace{-0.3cm}\includegraphics[width=4.0cm,height=4.0cm]{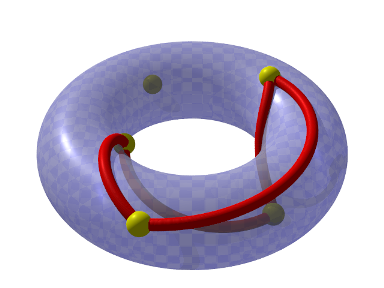}\hspace{-0.3cm}\includegraphics[width=4.0cm,height=4.0cm]{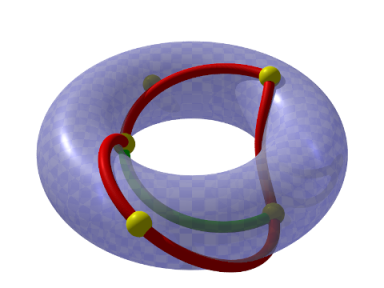}
\\
\vspace{-0.2cm}
\hspace{-0.3cm}\includegraphics[width=4.0cm,height=4.0cm]{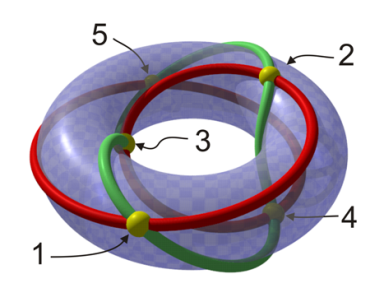}
\caption{\small{On the left and right upper panel two possible paths, corresponding to Y-type junctions, are illustrated on the surface of a torus representing the configuration space $(\phi,\psi)$. On the left and right middle panel two possible paths, corresponding to stable (left panel) and unstable (right panel) X-type junctions, are illustrated. The green line on the right middle panel represents a possible decay channel of the unstable X-type junction into two stable Y-type ones. On the lower panel the green and red paths (lighter and darker grey in black and white) illustrate the trajectories with constant $\chi$ and $\theta$, respectively.}}
\end{figure}

Consider a planar static domain wall perpendicular to the $z$ direction and assume that $\theta=\theta(z)$ and 
$\cos (\chi/2)= 0$. The only non-trivial equation of motion is given by
\begin{equation}
\label{eq:mot}
\frac{1}{10}\left(\frac{d\theta}{dz}\right)^2=2 V_\epsilon\cos^2 (\theta/2)\,,
\end{equation}
or equivalently
\begin{equation}
\frac{d\theta}{\cos (\theta/2)}=\pm {\sqrt {20 V_\epsilon}} dz \,,
\end{equation}
which has the solution
\begin{equation}
\label{eq:tan}
\tan^2\left(\frac{\pi+\theta}{4}\right)=e^{\pm{\sqrt {20 V_\epsilon}} z}\,,
\end{equation}
for a domain wall located at $z=0$.
Using Eq.~(\ref {eq:tan}) it is straightforward to show that
\begin{equation}
\cot \left(\frac{\pi+\theta}{2}\right)=\mp \sinh\left(\frac{z}{\delta_w}\right)\,,
\end{equation}
where $\delta_w=1/{\sqrt {20 V_\epsilon}}$ is the domain wall tickness.
In the following we shall drop the $\mp$ sign. It will be suficient to realize that for each solution $\theta=\theta(z)$, there will also be another solution given by $\theta=\theta(-z)$. 

The energy density, $\rho$, associated with the domain wall is 
\begin{eqnarray}
\rho(z) &=& \frac{1}{10}\left(\frac{d\theta}{dz}\right)^2 + 2 V_\epsilon\cos^2 (\theta/2) \nonumber\\
&=& 4 V_\epsilon\cos^2 (\theta/2)\,,
\end{eqnarray}
where Eq.~(\ref {eq:mot}) was used to obtain the final result. 
Finally the domain wall tension associated with a simple domain wall trajectory is given by
\begin{equation}
\sigma=\int_{-\infty}^{+\infty} \rho(z) dz  = 8 V_\epsilon \delta_w\,.
\end{equation}
In the case of a compound wall, in which both $\theta$ and $\chi$ vary along the wall, the corresponding tension is twice that 
of a simple domain wall
\begin{equation}
\sigma_{II}=2\sigma\,.
\end{equation}

In Carter's pentahedral model there is a simple domain wall trajectory between any of the five minima of the 
potential, with either constant $\theta$ or $\chi$. For  constant $\theta$ there is a simple domain wall trajectory between any two adjacent minima, with phases $(\phi,\psi)$, in the sequence
\begin{eqnarray}
1.\ (7\pi/5,-9\pi/5) \ 2. \ (\pi,-\pi) \ 3. \ (3\pi/5,-\pi/5) \nonumber \\ 
4.\ (\pi/5,3\pi/5)\  5.\  (-\pi/5,7\pi/5) \ 1. \ (-3\pi/5,11\pi/5)\,.
\end{eqnarray}
Here $\phi$ and $\psi$ vary in five successive steps of $\mp 2\pi/5$ and $\pm 4\pi/5$ respectively thus maintaining $\theta={\rm constant}$ (note that the phases are defined up to a multiple of $2\pi$). These trajectories are illustrated on the the lower panel of Fig.~1 by the red path (darker grey in black and white) on the surface of a torus with line element
\begin{equation}
dl^2=R_1^2 d \phi^2+(R_1 \cos \phi+R_2)^2 d \psi^2 
\end{equation}
where $R_1 < R_2$, representing the configuration space $(\phi,\psi)$. 

\begin{figure}[t!]
\hspace{-0.3cm}\includegraphics[width=4.5cm,height=4.5cm]{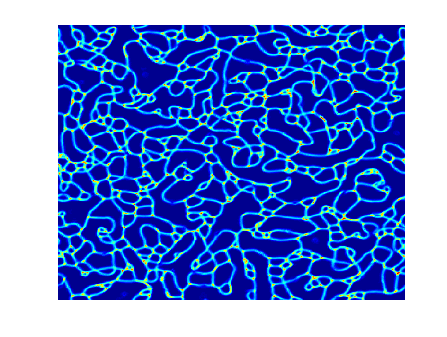}\hspace{-0.3cm}\includegraphics[width=4.5cm,height=4.5cm]{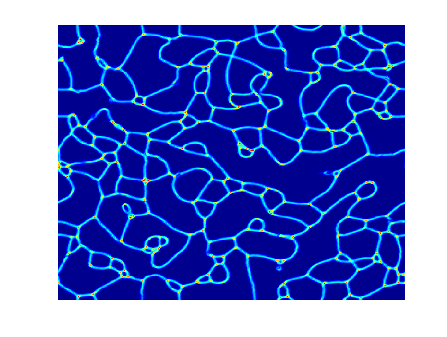}
\\
\vspace{-0.2cm}
\hspace{-0.3cm}\includegraphics[width=4.5cm,height=4.5cm]{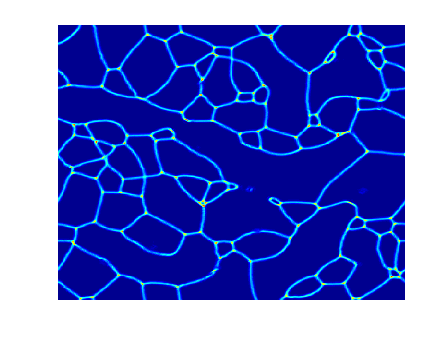}\hspace{-0.3cm}\includegraphics[width=4.5cm,height=4.5cm]{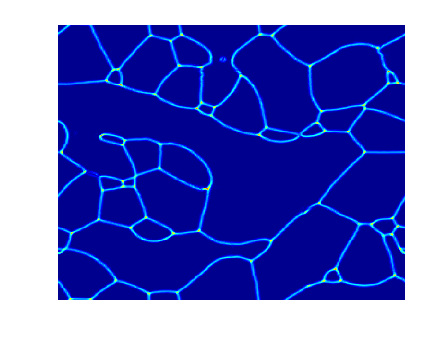}
\caption{\small{Matter-era evolution of a realization of Carter's pentahedral model with $\epsilon=0.2$. 
The simulation starts with random initial conditions with all the five minima having identical probability. 
Note that Y-type junctions are much more frequent than X-type ones. From left to right and top to
bottom, the horizon is approximately 1/10, 1/8, 1/6 and 1/4 of the box size, respectively.}}
\label{random0p2} 
\end{figure}

If $\chi$ is a constant then there is a simple domain wall trajectory between any two adjacent minima in the sequence
\begin{eqnarray}
1.\ (-13\pi/5,-9\pi/5) \ 4.\  (-9\pi/5,-7\pi/5) \ 2.\  (-\pi,-\pi)  \nonumber \\ 
5.\  (-\pi/5,-3\pi/5) \ 3.\  (3\pi/5,-\pi/5) \ 1.\  (7\pi/5,\pi/5)\,.
\end{eqnarray}
In this case $\phi$ and $\psi$ vary by successive steps of $\pm 4\pi/5$ and $\pm 2\pi/5$ respectively thus maintaining $\chi={\rm constant}$. These trajectories are illustrated by the green path (lighter grey in black and white) on the lower panel of Fig.~1.

There are Y-type junctions connecting three simple domain walls trajectories. Surrounding a Y-type junction there are two domain walls with constant $\theta$ (or $\chi$) and another one with constant $\chi$ (or $\theta$). A simple example is the configuration
\begin{eqnarray}
1.\ (-3\pi/5,\pi/5)\  4. \ (\pi/5,3\pi/5) \nonumber \\ 
3.\  (3\pi/5,-\pi/5) \ 1.\  (7\pi/5,\pi/5)\,,
\end{eqnarray}
which corresponds to two domain walls with constant $\chi$ ($1-4$ and $1-3$) and one with constant 
$\theta$ ($3-4$). The above trajectory is illustrated by the red path on the left upper panel of Fig.~1. 
The overall change in the phase $\phi$ is equal to $2 \pi$. In fact there must always be jump of $2\pi$ in either $\phi$ or $\psi$ around a Y-type junction.

\begin{figure}[t!]
\hspace{-0.3cm}\includegraphics[width=4.5cm,height=4.5cm]{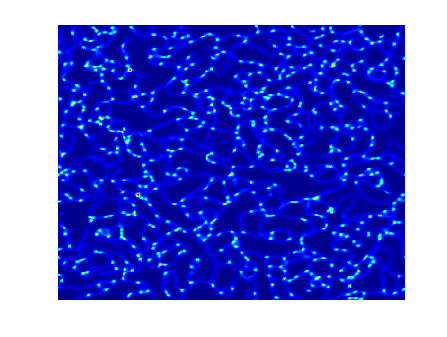}\hspace{-0.3cm}\includegraphics[width=4.5cm,height=4.5cm]{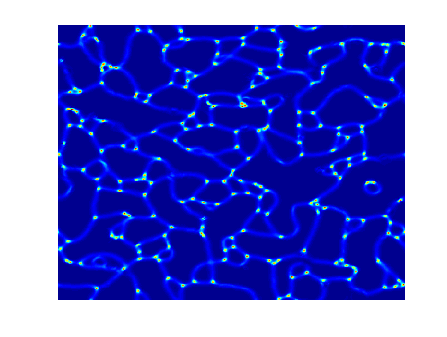}
\\
\vspace{-0.2cm}
\hspace{-0.3cm}\includegraphics[width=4.5cm,height=4.5cm]{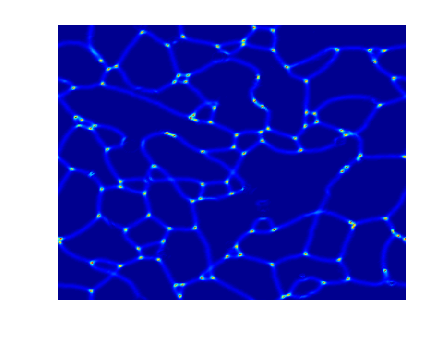}\hspace{-0.3cm}\includegraphics[width=4.5cm,height=4.5cm]{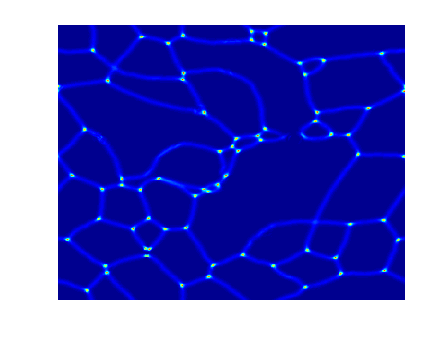}
\caption{\small{Same as Fig.~1, except that now $\epsilon=0.05$.}}
\end{figure}

Another example corresponding to a Y-type junction is the configuration
\begin{eqnarray}
1.\ (-3\pi/5,\pi/5)\  2. \ (-\pi,\pi) \nonumber \\ 
5.\  (-\pi/5,7\pi/5) \ 1.\  (-3\pi/5,11\pi/5)\,,
\end{eqnarray}
where two domain walls with constant $\theta$ ($1-2$ and $1-5$) and one with constant 
$\chi$ ($2-5$) meet.  In this case it is the overall change in $\psi$ that is equal to $2 \pi$. This trajectory is illustrated by the red path on the right upper panel of Fig.~1.

What about X-type junctions ? Is there a trajectory in which $\phi$ and $\psi$ are continuous around a X-type junction ? The answer is yes. For example, both $\phi$ and $\psi$ can be made continuous around the X-type junction described by the following configuration 
\begin{eqnarray}
1.\  (7\pi/5,\pi/5)\  2.\  (\pi,\pi)\  4.\  (\pi/5,3\pi/5)\nonumber\\ 
3.\  (3\pi/5,-\pi/5)\ 1.\  (7\pi/5,\pi/5)\,.
\end{eqnarray}
This trajectory is illustrated by the red path on the left middle panel of Fig.~1. 
As correctly pointed out in \cite{Carter:2004dk,Carter:2006cf}, Carter's penthedral model allows for square domain wall lattice solutions that are stable, if $\epsilon$ is sufficiently small. However, as we will show in the following section, such lattices are never generated from realistic initial conditions.

Around a X-type junction where three walls with constant $\theta$ (or $\chi$)  meet one wall with constant $\chi$ (or $\theta$), 
both $\phi$ and $\psi$ must change by a factor of $2 \pi$. Consider the following example which is 
illustrated by the red path on the right middle panel of Fig.~1
\begin{eqnarray}
1.\ (-3\pi/5,-9\pi/5) \ 2. \ (-\pi,-\pi) \ 5. \ (-\pi/5,-3\pi/5) \nonumber \\ 
3.\ (3\pi/5,-\pi/5)\  1. \ (7\pi/5,\pi/5)\,.
\end{eqnarray}
In this case the energy of the junction associated with the presence of a string is greater, by a factor of 2, compared to Y-type junctions. Hence, the string does nothing for the stability of the junction. Such X-type junction would be unstable and decay into a pair of Y-type junctions, even if  $\epsilon$ is small (the green line represents a possible decay channel). This is the reason why, in the context of Carter's pentahedral model, Y-type junctions are preferred, with exception of very specific realizations.

\section{Simulations}

In order to test our analytical expectations we will now present the results of a few $256^2$ simulations in two spatial dimensions. Although these simulations are relatively small in size and dynamical range, they are more than enough to support our analysis. In all the simulations we use the PRS algorithm \cite{Press} modifying the domain wall thickness in order to ensure a fixed comoving resolution. More details about the numerical code can be found in  \cite{Avelino:2008ve} and references therein.

\begin{figure}[t!]
\includegraphics[width=9.6cm,height=3.0cm]{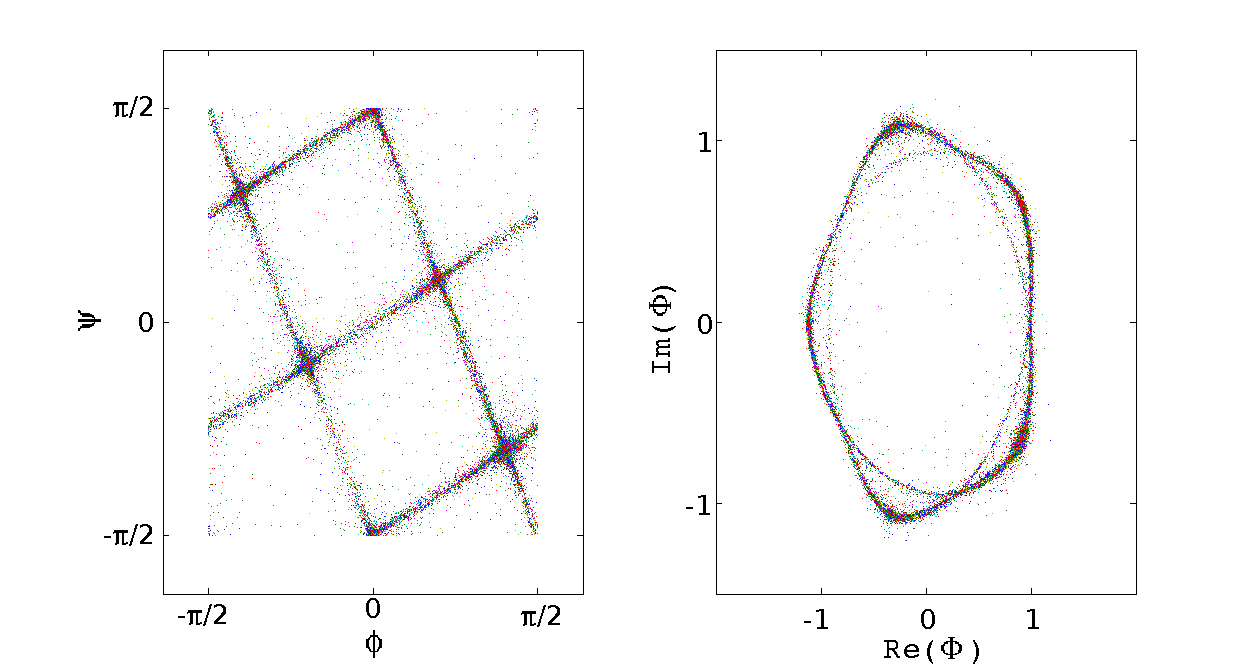}
\caption{\small{The configuration space distribution for the last time step of the
simulation in Fig.~2. On the left panel the $x$ and $y$ axis represent the phases $\phi$ and $\psi$, respectively.
On the right panel the $x$ and $y$ axis represent ${\rm Re(\Phi)}=|\Phi|\cos\phi$ and ${\rm Im(\Phi)}=|\Phi|\sin\phi$.}}
\label{Cartefase0p2} 
\end{figure}

\begin{figure}[t!]
\includegraphics[width=9.6cm,height=3.0cm]{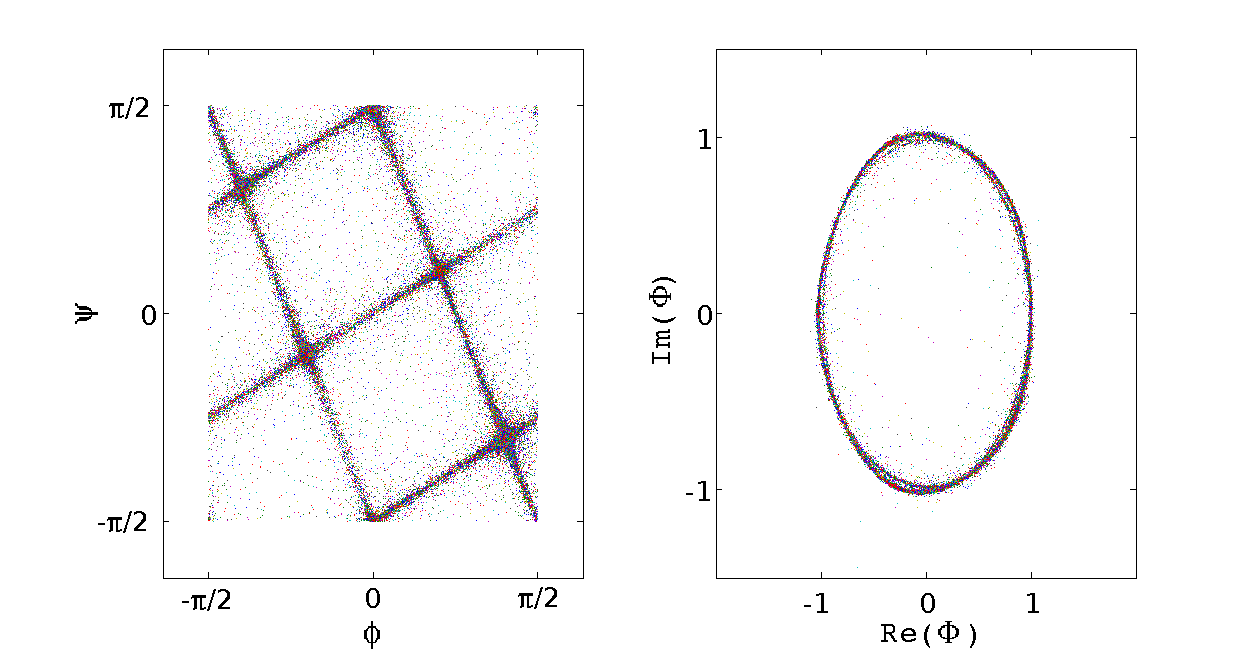}
\caption{\small{The same as in Fig. \protect \ref{Cartefase0p2} for $\mathbf{\epsilon}=0.05$.}}
\label{Cartefase0p05} 
\end{figure}

Fig.~2 shows four snapshots of a matter era simulation of a realization of Carter's pentahedral model ($\epsilon=0.2$) with random initial conditions. At each grid point, one of the minima was randomly assigned, all the minima having equal probability. The cosmic time, $t$, is increasing from left to right and top to bottom (the horizon is approximately 1/10, 1/8, 1/6 and 1/4 of the box size respectively). The simulations show that Y-type junctions are much more frequent than X-type ones. 
This is not surprising since the probability that the combination of two Y-type junctions will give rise to one stable X-type can be easily calculated and is equal to $2/9$, assuming that the corresponding minima are randomly chosen with equal probability, subject to the constraint the same minima cannot be assigned to both sides of a domain wall. On the other hand, the probability that the collapse of a square domain with Y-type junctions at the vertices will give rise to a stable X-type junction is equal to $1/21$, 
again assuming a random configuration.  Furthermore,  this does not take into consideration that stable X-type junctions may break into two Y-type junctions if enough energy is available. However, some rare stable X-type junctions can still be identified in 
the simulations.

\begin{figure}[t!]
\hspace{-0.3cm}\includegraphics[width=4.5cm,height=4.5cm]{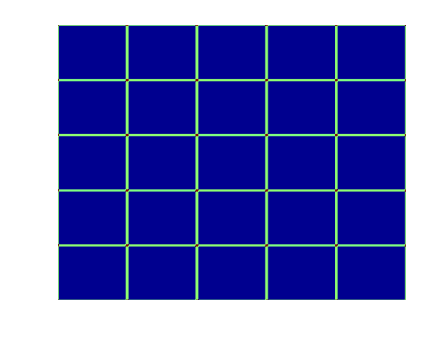}\hspace{-0.3cm}\includegraphics[width=4.5cm,height=4.5cm]{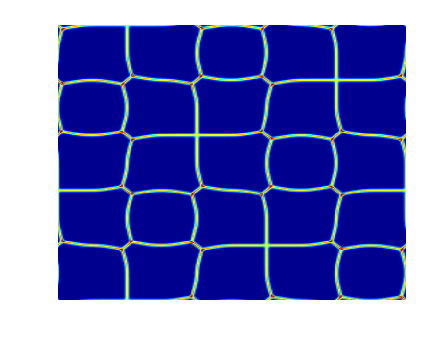}
\\
\vspace{-0.2cm}
\hspace{-0.3cm}\includegraphics[width=4.5cm,height=4.5cm]{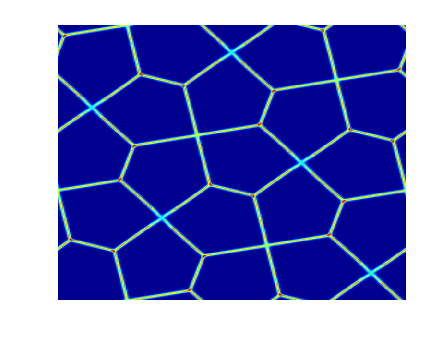}\hspace{-0.3cm}\includegraphics[width=4.5cm,height=4.5cm]{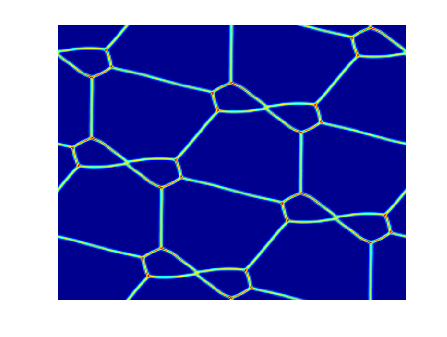}
\caption{\small{Evolution of a periodic square lattice of Carter's pentahedral model.
The initial configuration of minima was chosen to allow for both stable and unstable 
X-type junctions.}}
\end{figure}

Fig.~3 is similar to Fig~2 except that now $\epsilon=0.05$. As a consequence, the energy density inside the domain walls 
is reduced by a factor of $4$ while their thickness increases by a factor of $2$. On the other hand, the strings remain roughly the same. Of course, in the $\epsilon \to 0$ limit the dynamics would be completely dominated by the 
strings. However, in this limit the thickness of the domain walls becomes 
very large ($\delta_w \propto \epsilon^{-1/2}$) and the domain wall network would no longer be well defined. In any case, this would not help domain walls as a possible dark energy candidate since, in that case, the contribution of the junctions to the energy density would be the dominant one, thus leading to an equation of state parameter significantly greater than $-2/3$.  
Moreover, the strings have a small impact on the overall dynamics as long as the average domain wall energy density dominates over that associated with the junctions. This happens for $\sigma L \gg \mu \sim 1$ or equivalently $\delta_w/L \ll 1$, where $L$ is the characteristic scale of the network. Such condition is always verified as long as the thickness of the domain walls is much smaller their typical curvature scale. In fact, the string energy per unit length ($\mu \sim 1$) of a stable Y-type junction is of the same order as the energy per unit length of a stable X-type junction ($\sim V_\epsilon \delta_w^2 \sim 1$), which means that 
X-type junctions are configurations of delicate equilibrium, susceptible to decay in the presence of relatively small 
perturbations. Hence, even the rare stable X-type junctions which appear in the simulations would probably not be there if the domain wall thickness had not to be artificially enlarged in order to ensure that the domain walls were resolved by the numerical code.

Fig.~4 shows the configuration space distribution for the last time step of the simulation in Fig.~2. 
On the left panel $x$ and $y$ axis represent $\phi$ and $\psi$, respectively. The left panel shows that only simple domain wall trajectories with constant $\theta=2\phi+\psi$ (or $\chi=2\psi -\phi$) appear in the simulations. On the right panel, the $x$ and $y$ axis represent ${\rm Re(\Phi)}=|\Phi|\cos\phi$ and ${\rm Im(\Phi)}=|\Phi|\sin\phi$, respectively. The five different minima corresponding to a constant value of $|\Phi|=\sqrt{1/\left(1-\epsilon\right)}\simeq1.11$, as well as the corresponding domain wall 
trajectories in $\Phi$, can be easily identified. Fig.~5 is similar to Fig~4 except that now $\epsilon=0.05$ so that the minima on the left correspond to $|\Phi|\simeq1.01$. As a result, the domain wall trajectories appear as nearly circular orbits on the left panel of Fig.~5.

Fig.~6 shows the evolution of a hand-made periodic square lattice realization of Carter's pentahedral model with $\epsilon=0.2$.
The initial configuration of minima was chosen to allow for X-type junctions corresponding to a continuous $\phi$ and $\psi$, and X-type junctions around which both $\phi$ and $\psi$ change by a factor 
of $2 \pi$. As expected, the simulations show that the former are stable while the later are unstable and decay into two stable Y-type junctions.

It is also possible to choose the initial conditions in a way that a square lattice with only X-type junctions is formed and we have verified that such a configuration is stable, as claimed by Carter \cite{Carter:2004dk,Carter:2006cf}. However, one should bear in mind that it corresponds to a very specific set of initial conditions which would violate causality, if they were to extend over scales larger than the particle horizon.

\section{Conclusions}

In this paper we confirmed that there are very special realizations of Carter's pentahedral 
model, corresponding to square lattice configurations with X-type junctions, which could be 
stable. However, we have shown that more realistic realizations of Carter's pentahedral model, such as 
those with random initial conditions, give rise to a network with Y-type junctions. 
This leads to a domain wall network whose properties are virtually indistinguishable 
from those of a specific realization of the ideal class of models with 4 real scalar fields (and 5 
minima), with similar initial conditions. The ideal class of models has been studied in detail in 
\cite{Avelino:2006xf,Avelino:2008ve} where a compelling evidence for a gradual 
approach to scaling, with $L \propto t$, was found both in the radiation and matter eras. 
As a result, and in spite of its very interesting topological properties, Carter's 
pentahedral model does not naturally lead to a frustrated network with $v \sim 0$ and 
$L \ll t$, a necessary 
condition for domain walls to provide a contribution to the dark energy budget.
There are other models which allow for X-type junctions (see for example, 
\cite{Kubotani:1991kw,Bazeia:2005wt}) but they also do not lead to a frozen network, 
starting from random initial conditions \cite{Avelino:2006xf,Avelino:2008ve}.

%%%%%%%%%%%%%%%%%%%%%%%%%%%%%%%%%%%%%%%%%%%%%%%%%%%%%
\begin{acknowledgments}

We thank Carlos Herdeiro and Lara Sousa for useful discussions. This work is part of a collaboration between Departamento de F\'\i sica, Universidade Federal da Para\'\i ba, Brazil, and Departamento de F\'\i sica, Universidade do Porto, Portugal, supported by the CAPES-GRICES project.This work was also funded by FCT (Portugal) through contract CERN/FP/83508/2008.

\end{acknowledgments}

%%%%%%%%%%%%%%%%%%%%%%%%%%%%%%%%%%%%%%%%%%%%%%%%%%%%%%%%%%

\bibliography{Carter}

\begin{thebibliography}{16}
\expandafter\ifx\csname natexlab\endcsname\relax\def\natexlab#1{#1}\fi
\expandafter\ifx\csname bibnamefont\endcsname\relax
  \def\bibnamefont#1{#1}\fi
\expandafter\ifx\csname bibfnamefont\endcsname\relax
  \def\bibfnamefont#1{#1}\fi
\expandafter\ifx\csname citenamefont\endcsname\relax
  \def\citenamefont#1{#1}\fi
\expandafter\ifx\csname url\endcsname\relax
  \def\url#1{\texttt{#1}}\fi
\expandafter\ifx\csname urlprefix\endcsname\relax\def\urlprefix{URL }\fi
\providecommand{\bibinfo}[2]{#2}
\providecommand{\eprint}[2][]{\url{#2}}

\bibitem[{\citenamefont{Komatsu et~al.}(2009)}]{Komatsu:2008hk}
\bibinfo{author}{\bibfnamefont{E.}~\bibnamefont{Komatsu}} \bibnamefont{et~al.}
  (\bibinfo{collaboration}{WMAP}), \bibinfo{journal}{Astrophys. J. Suppl.}
  \textbf{\bibinfo{volume}{180}}, \bibinfo{pages}{330} (\bibinfo{year}{2009}),
  \eprint{0803.0547}.

\bibitem[{\citenamefont{Frieman et~al.}(2008)\citenamefont{Frieman, Turner, and
  Huterer}}]{Frieman:2008sn}
\bibinfo{author}{\bibfnamefont{J.}~\bibnamefont{Frieman}},
  \bibinfo{author}{\bibfnamefont{M.}~\bibnamefont{Turner}}, \bibnamefont{and}
  \bibinfo{author}{\bibfnamefont{D.}~\bibnamefont{Huterer}},
  \bibinfo{journal}{Ann. Rev. Astron. Astrophys.}
  \textbf{\bibinfo{volume}{46}}, \bibinfo{pages}{385} (\bibinfo{year}{2008}),
  \eprint{0803.0982}.

\bibitem[{\citenamefont{Copeland et~al.}(2006)\citenamefont{Copeland, Sami, and
  Tsujikawa}}]{Copeland:2006wr}
\bibinfo{author}{\bibfnamefont{E.~J.} \bibnamefont{Copeland}},
  \bibinfo{author}{\bibfnamefont{M.}~\bibnamefont{Sami}}, \bibnamefont{and}
  \bibinfo{author}{\bibfnamefont{S.}~\bibnamefont{Tsujikawa}},
  \bibinfo{journal}{Int. J. Mod. Phys.} \textbf{\bibinfo{volume}{D15}},
  \bibinfo{pages}{1753} (\bibinfo{year}{2006}), \eprint{hep-th/0603057}.

\bibitem[{\citenamefont{Bucher and Spergel}(1999)}]{Bucher:1998mh}
\bibinfo{author}{\bibfnamefont{M.}~\bibnamefont{Bucher}} \bibnamefont{and}
  \bibinfo{author}{\bibfnamefont{D.~N.} \bibnamefont{Spergel}},
  \bibinfo{journal}{Phys. Rev.} \textbf{\bibinfo{volume}{D60}},
  \bibinfo{pages}{043505} (\bibinfo{year}{1999}), \eprint{astro-ph/9812022}.

\bibitem[{\citenamefont{Avelino
  et~al.}(2006{\natexlab{a}})\citenamefont{Avelino, Martins, Menezes, Menezes,
  and Oliveira}}]{PinaAvelino:2006ia}
\bibinfo{author}{\bibfnamefont{P.~P.} \bibnamefont{Avelino}},
  \bibinfo{author}{\bibfnamefont{C.~J. A.~P.} \bibnamefont{Martins}},
  \bibinfo{author}{\bibfnamefont{J.}~\bibnamefont{Menezes}},
  \bibinfo{author}{\bibfnamefont{R.}~\bibnamefont{Menezes}}, \bibnamefont{and}
  \bibinfo{author}{\bibfnamefont{J.~C. R.~E.} \bibnamefont{Oliveira}},
  \bibinfo{journal}{Phys. Rev.} \textbf{\bibinfo{volume}{D73}},
  \bibinfo{pages}{123519} (\bibinfo{year}{2006}{\natexlab{a}}),
  \eprint{astro-ph/0602540}.

\bibitem[{\citenamefont{Avelino
  et~al.}(2006{\natexlab{b}})\citenamefont{Avelino, Martins, Menezes, Menezes,
  and Oliveira}}]{Avelino:2006xy}
\bibinfo{author}{\bibfnamefont{P.~P.} \bibnamefont{Avelino}},
  \bibinfo{author}{\bibfnamefont{C.~J. A.~P.} \bibnamefont{Martins}},
  \bibinfo{author}{\bibfnamefont{J.}~\bibnamefont{Menezes}},
  \bibinfo{author}{\bibfnamefont{R.}~\bibnamefont{Menezes}}, \bibnamefont{and}
  \bibinfo{author}{\bibfnamefont{J.~C. R.~E.} \bibnamefont{Oliveira}},
  \bibinfo{journal}{Phys. Rev.} \textbf{\bibinfo{volume}{D73}},
  \bibinfo{pages}{123520} (\bibinfo{year}{2006}{\natexlab{b}}),
  \eprint{hep-ph/0604250}.

\bibitem[{\citenamefont{Avelino et~al.}(2007)\citenamefont{Avelino, Martins,
  Menezes, Menezes, and Oliveira}}]{Avelino:2006xf}
\bibinfo{author}{\bibfnamefont{P.~P.} \bibnamefont{Avelino}},
  \bibinfo{author}{\bibfnamefont{C.~J. A.~P.} \bibnamefont{Martins}},
  \bibinfo{author}{\bibfnamefont{J.}~\bibnamefont{Menezes}},
  \bibinfo{author}{\bibfnamefont{R.}~\bibnamefont{Menezes}}, \bibnamefont{and}
  \bibinfo{author}{\bibfnamefont{J.~C. R.~E.} \bibnamefont{Oliveira}},
  \bibinfo{journal}{Phys. Lett.} \textbf{\bibinfo{volume}{B647}},
  \bibinfo{pages}{63} (\bibinfo{year}{2007}), \eprint{astro-ph/0612444}.

\bibitem[{\citenamefont{Battye and Moss}(2006)}]{Battye:2006pf}
\bibinfo{author}{\bibfnamefont{R.~A.} \bibnamefont{Battye}} \bibnamefont{and}
  \bibinfo{author}{\bibfnamefont{A.}~\bibnamefont{Moss}},
  \bibinfo{journal}{Phys. Rev.} \textbf{\bibinfo{volume}{D74}},
  \bibinfo{pages}{023528} (\bibinfo{year}{2006}), \eprint{hep-th/0605057}.

\bibitem[{\citenamefont{Avelino et~al.}(2008)\citenamefont{Avelino, Martins,
  Menezes, Menezes, and Oliveira}}]{Avelino:2008ve}
\bibinfo{author}{\bibfnamefont{P.~P.} \bibnamefont{Avelino}},
  \bibinfo{author}{\bibfnamefont{C.~J. A.~P.} \bibnamefont{Martins}},
  \bibinfo{author}{\bibfnamefont{J.}~\bibnamefont{Menezes}},
  \bibinfo{author}{\bibfnamefont{R.}~\bibnamefont{Menezes}}, \bibnamefont{and}
  \bibinfo{author}{\bibfnamefont{J.~C. R.~E.} \bibnamefont{Oliveira}},
  \bibinfo{journal}{Phys. Rev.} \textbf{\bibinfo{volume}{D78}},
  \bibinfo{pages}{103508} (\bibinfo{year}{2008}), \eprint{0807.4442}.

\bibitem[{\citenamefont{Carter}(2005)}]{Carter:2004dk}
\bibinfo{author}{\bibfnamefont{B.}~\bibnamefont{Carter}},
  \bibinfo{journal}{Int. J. Theor. Phys.} \textbf{\bibinfo{volume}{44}},
  \bibinfo{pages}{1729} (\bibinfo{year}{2005}), \eprint{hep-ph/0412397}.

\bibitem[{\citenamefont{Battye et~al.}(2005)\citenamefont{Battye, Carter,
  Chachoua, and Moss}}]{Battye:2005hw}
\bibinfo{author}{\bibfnamefont{R.~A.} \bibnamefont{Battye}},
  \bibinfo{author}{\bibfnamefont{B.}~\bibnamefont{Carter}},
  \bibinfo{author}{\bibfnamefont{E.}~\bibnamefont{Chachoua}}, \bibnamefont{and}
  \bibinfo{author}{\bibfnamefont{A.}~\bibnamefont{Moss}},
  \bibinfo{journal}{Phys. Rev.} \textbf{\bibinfo{volume}{D72}},
  \bibinfo{pages}{023503} (\bibinfo{year}{2005}), \eprint{hep-th/0501244}.

\bibitem[{\citenamefont{Battye et~al.}(2006)\citenamefont{Battye, Chachoua, and
  Moss}}]{Battye:2005ik}
\bibinfo{author}{\bibfnamefont{R.~A.} \bibnamefont{Battye}},
  \bibinfo{author}{\bibfnamefont{E.}~\bibnamefont{Chachoua}}, \bibnamefont{and}
  \bibinfo{author}{\bibfnamefont{A.}~\bibnamefont{Moss}},
  \bibinfo{journal}{Phys. Rev.} \textbf{\bibinfo{volume}{D73}},
  \bibinfo{pages}{123528} (\bibinfo{year}{2006}), \eprint{hep-th/0512207}.

\bibitem[{\citenamefont{Carter}(2008)}]{Carter:2006cf}
\bibinfo{author}{\bibfnamefont{B.}~\bibnamefont{Carter}},
  \bibinfo{journal}{Class. Quant. Grav.} \textbf{\bibinfo{volume}{25}},
  \bibinfo{pages}{154001} (\bibinfo{year}{2008}), \eprint{hep-ph/0605029}.

\bibitem[{\citenamefont{Press et~al.}(1989)\citenamefont{Press, Ryden, and
  Spergel}}]{Press}
\bibinfo{author}{\bibfnamefont{W.~H.} \bibnamefont{Press}},
  \bibinfo{author}{\bibfnamefont{B.~S.} \bibnamefont{Ryden}}, \bibnamefont{and}
  \bibinfo{author}{\bibfnamefont{D.~N.} \bibnamefont{Spergel}},
  \bibinfo{journal}{Astrophys. J.} \textbf{\bibinfo{volume}{347}},
  \bibinfo{pages}{590} (\bibinfo{year}{1989}).

\bibitem[{\citenamefont{Kubotani}(1992)}]{Kubotani:1991kw}
\bibinfo{author}{\bibfnamefont{H.}~\bibnamefont{Kubotani}},
  \bibinfo{journal}{Prog. Theor. Phys.} \textbf{\bibinfo{volume}{87}},
  \bibinfo{pages}{387} (\bibinfo{year}{1992}).

\bibitem[{\citenamefont{Bazeia et~al.}(2006)\citenamefont{Bazeia, Brito, and
  Losano}}]{Bazeia:2005wt}
\bibinfo{author}{\bibfnamefont{D.}~\bibnamefont{Bazeia}},
  \bibinfo{author}{\bibfnamefont{F.~A.} \bibnamefont{Brito}}, \bibnamefont{and}
  \bibinfo{author}{\bibfnamefont{L.}~\bibnamefont{Losano}},
  \bibinfo{journal}{Europhys. Lett.} \textbf{\bibinfo{volume}{D76}},
  \bibinfo{pages}{374} (\bibinfo{year}{2006}), \eprint{hep-th/0512331}.

\end{thebibliography}

\end{document}